\documentclass[a4paper]{jpconf}
\usepackage{amsmath,amssymb,graphicx,bm}
\usepackage{color}
\usepackage[bottom]{footmisc}

\newcommand{\beq}{\begin{equation}}
\newcommand{\eeq}{\end{equation}}
\newcommand{\bea}{\begin{eqnarray}}
\newcommand{\eea}{\end{eqnarray}}
\newcommand{\ba}{\begin{align}}
\newcommand{\ea}{\end{align}}
\newcommand{\bfig}{\begin{figure}}
\newcommand{\efig}{\end{figure}}

\newcommand{\gev}{\, \text{GeV}}

\newcommand{\tin}{t_{\rm in}}

\newcommand{\tplus}{t_{+}}

\begin{document}
\title{The $K\pi$ form factors from Analyticity and Unitarity}

\author{Gauhar Abbas$^{a,\star}$, B.Ananthanarayan$^{a,\ast}$, I.Caprini$^{b,\dagger}$, 
I.Sentitemsu Imsong$^{a,\ddagger}$}

\address{$^a$ Centre for High Energy Physics,
Indian Institute of 
Science, Bangalore 560 012, India \\
$^b$ Horia Hulubei National Institute for Physics and Nuclear Engineering,
P.O.B. MG-6, 077125 Magurele, Romania}

\ead{$^{\star}$gabbas@cts.iisc.ernet.in \\
\hspace{1.2cm}$^{\ast}$anant@cts.iisc.ernet.in \\
\hspace{1.2cm}$^{\dagger}$caprini@theory.nipne.ro \\
\hspace{1.2cm}$^{\ddagger}$senti@cts.iisc.ernet.in}

\begin{abstract}
Analyticity and unitarity techniques are employed to obtain bounds on the  shape
 parameters 
of the scalar and vector
form factors of semileptonic $K_{l3}$ decays.  For this purpose we use
vector and scalar  correlators evaluated 
in pQCD, a low energy theorem for scalar form factor, lattice results for 
the ratio of kaon and pion decay constants, chiral perturbation theory 
calculations for the scalar form factor at the Callan-Treiman point and 
experimental information on the phase and modulus of $K\pi $ form factors up 
to an energy 
$\tin=1 {\rm GeV}^2$.
We further derive regions on the real axis and in the complex-energy 
plane where the form factors 
cannot have zeros. 

\end{abstract}

\section{Introduction} \label{sec:intro}
\bigskip

$K_{l3}$ decays are important for determining the 
matrix element $V_{us}$, for recent reviews, see  \cite{Antonelli:2010yf,Cirigliano:2011ny} of
the Cabibbo-Kobayashi-Maskawa (CKM) which in turn is crucial for testing the
unitarity of the CKM matrix. Information on the experimental sector is rich,
in particular for the $K_{\ell3}$ decay rates, which were measured 
by a number of experiments BNL-E865 \cite{Sher:2003fb}, KLOE \cite{Ambrosino:2006gn},  
ISTRA+ \cite{Romanovsky:2007qb}, KTEV \cite{Abouzaid:2009ry}
and more recently a new analysis by NA48 \cite{Veltri:2011zk}. Recent lattice 
studies 
have also been carried out, see refs.\cite{Lellouch:2009fg,Durr:2010hr,
Colangelo:2010et,Ramos:2011wp}.

The decay of a kaon to a pion, a charged lepton and a neutrino
is described by the matrix element
\begin{equation}
\langle \pi^0(p') | \overline{s}\gamma_\mu u |K^+(p) \rangle  =  
\frac{1}{ \sqrt{2}}[(p'+p)_\mu f_+(t)+(p-p')_\mu f_-(t)],
\end{equation}
where $f_+(t)$ is the vector form factor and the combination
\begin{equation}\label{eq:f0}
f_0(t)=f_+(t)+\frac{t}{M_K^2-M_\pi^2} f_-(t)
\end{equation}
is known as the scalar form factor. The expansion  at $t=0$
\begin{equation}
       f_k(t) = f_+(0)\left(1 +\lambda_k'\, \frac{t}{M_\pi^2} +  \frac{1}{2} 
\lambda_k''\, \frac{t^2}{M_\pi^4} + \cdots\right)
       \label{eq:f0exp},
\end{equation}
defines the slope $\lambda_k'$ and the curvature $\lambda_k''$ parameters where $k=0$ 
denotes
the scalar and $k=+$ denotes the vector form factor.  
The precise determination of the element $|V_{us}|$ depends on how accurate the
parametrization of the
form factor is.
To improve the precision and to provide bounds on the shape parameters of the 
form factors, 
we use inputs 
coming from 
certain low energy theorems,  perturbative QCD, lattice computations  and 
chiral perturbation theory. Our techniques allow us to 
incorporate the phase and 
modulus information of the form factors. 
We also apply the technique to find regions on
the real axis and in
the complex $t$-plane where zeros are excluded.
The knowledge of zeros is 
of interest, for instance, for the dispersive methods  
(Omn\`es-type representations) and 
for testing specific models of the form factors. For more details, see refs. 
\cite{Abbas:2010ns,Abbas:2010jc,Abbas:2009dz,Abbas:2011qj}.

\section{Formalism}\label{formalism}

\bigskip
The formalism described in \cite{Abbas:2010ns,Abbas:2010jc,Abbas:2009dz} and
in the
contribution to these Proceedings \cite{anant} , 
exploits an integral of the
type
\beq
\int^{\infty}_{\tplus } dt\ \rho_{+,0}(t) |f_{+,0}(t)|^{2} \leq I_{+,0},
        \label{eq:I}
\eeq
along the unitarity cut, whose upper bound is known from a dispersion relation, 
satisfied by a certain QCD correlator. For the scalar form factor this reads
\begin{equation}
\label{chi}
\chi_{_0}(Q^2)\equiv \frac{\partial}{ \partial q^2} \left[ q^2\Pi_0 \right]
= \frac{1}{\pi}\int_{t_+}^\infty\!dt\, \frac{t {\rm Im}\Pi_0(t)}{ (t+Q^2)^2} \,,
\end{equation}

\begin{equation}
\label{im2}
{\rm Im} \Pi_0(t) \ge \frac{3}{2} \frac{t_+ t_-}{ 16\pi}
\frac{[(t-t_+)(t-t_-)]^{1/2}}{ t^3} |f_0(t)|^2  \,,
\end{equation}
with $t_\pm=(M_K \pm M_\pi)^2$.
Analogues expression, involving a suitable correlator denoted by $\chi_1(Q^2)$, can be
written down for  the vector form factor.  
We can now use the conformal map $t\to z(t)$
\begin{equation}\label{eq:z}
z(t)=\frac{\sqrt{t_+}-\sqrt{t_+-t}}{\sqrt{t_+}+\sqrt{t_+-t}}\,,
\end{equation}
that maps the cut $t$-plane onto the unit disc $|z|<1$ in the $z\equiv z(t)$ 
plane, with $\tplus$  mapped onto $z = 1$, the point at 
infinity to $z = -1$ and the origin to $z=0$. Using this map, we
cast eqn.(\ref{eq:I}) into a canonical form, incorporate phase and modulus information 
as well as the Callan-Treiman relations and finally employ a determinant 
 for 
obtaining bounds 
on the shape parameters and for finding regions of excluded zeros in the 
complex $t$-plane.

\section{Inputs}

\bigskip
The essential inputs of our formalism, the vector $\chi_1(Q^2)$ and the scalar 
$\chi_0(Q^2)$ correlators, can 
be calculated in perturbative QCD 
up to order $\alpha_s^4$ for $Q^2 >>\Lambda^2_{QCD}$ 
\cite{Baikov:2005rw,Baikov:2008jh}.
We get $\chi_1(2 \gev)=(343.8 \pm  51.6 )\times 10^{-5}
\, {\rm GeV}^{-2}$ and  $\chi_0(2 \gev)= (253\pm 68)\times 10^{-6}$,
see also \cite{Abbas:2010ns}.
An improvement can be achieved when we 
implement theoretical and experimental information into the formalism of unitarity 
bounds.
The first improvement comes when we use the the value of vector form factor at zero 
momentum
transfer. Recent determinations from the lattice give $f_+(0)=0.964(5)$ 
\cite{Boyle:2007qe}. 
We can also use two low energy theorems, namely soft pion and soft kaon theorems, for the 
improvement of 
the bounds on the slope and curvature 
parameters in the scalar case.
The soft pion theorem relates the value of
scalar form factor at first Callan-Treiman piont $\Delta_{K\pi}\equiv M_K^2-M_\pi^2$  to 
the ratio  $F_K/F_\pi$ of the decay constants 
\cite{Callan:1966hu,Dashen:1969bh}:
\beq\label{eq:CT1}
f_0(\Delta_{K\pi})=F_K /F_\pi  +\Delta_{CT}.
\eeq 
Recent lattice evaluations with $N_f$ = 2 + 1 flavors of sea quarks
give $F_K/F_\pi=1.193\pm 0.006$ \cite{Lellouch:2009fg,Durr:2010hr}.  
In the isospin limit, $\Delta_{CT}= -3.1\times 10^{-3}$ to one loop 
\cite{Gasser:1984ux} and
$\Delta_{CT}\simeq 0$ to two-loops in chiral perturbation theory 
\cite{Kastner:2008ch,
Bijnens:2003uy,Bijnens:2007xa}. 

At $\bar{\Delta}_{K\pi}(=-\Delta_{K\pi})$, a soft-kaon result 
\cite{Oehme} relates the 
value
of the scalar form factor to $F_\pi/F_K$ 
\beq\label{eq:CT2}
f_0(-\Delta_{K\pi})=F_\pi/F_K  +\bar{\Delta}_{CT}.
\eeq
A calculation  in ChPT to one-loop in the isospin limit \cite{Gasser:1984ux} gives 
$\bar{\Delta}_{CT}=0.03$, 
but the  higher order ChPT corrections are expected to be  larger in this case. 
As discussed in \cite{Abbas:2009dz}, due to the poor knowledge of $\bar{\Delta}_{CT}$, 
the low-energy theorem 
eqn.(\ref{eq:CT2}) is not useful for  further constraining the shape of the 
$K_{\ell 3}$ 
form factors 
at low energies.  On the other hand, we obtain from the same machinery, the stringent
bound on the quantity  $\bar{\Delta}_{CT}$ which is 
$-0.046 \leq \bar{\Delta}_{CT} \leq 0.014$.

Further improvement of the bounds can be achieved if the phase of the form factor 
along the elastic part of the unitarity cut is known from an independent source.  
In our calculations  we use below $\tin$ the phases from \cite{Buettiker:2003pp,
ElBennich:2009da} for the 
scalar form factor, and from \cite{Moussallam:2007qc, Bernard:2009zm} for the vector form 
factor. 
Above $\tin$ we take  $\delta(t)$ Lipschitz continuous, i.e., a smooth function 
approaching $\pi$ at high energies. 
The results are  independent of the choice of the phase for  $t>\tin$. 
We can further improve the bounds if the modulus of the form factor is known along the 
unitarity cut, $t \leq \tin$: we can shift the branch point from $t_{\pm}$ to $t_{in}$ 
by subtracting the low energy integral from the
integral Eq. (\ref{eq:I}).
In order to estimate  the low-energy integral, which is the value of the integral 
contribution from
$t_{+}$ to $t_{\rm in}$, see for expressions  ref.\cite{Abbas:2010jc}, 
we use the  Breit-Wigner parameterizations of  $|f_+(t)|$ and $|f_0(t)|$ in terms of 
the resonances given by the
Belle Collaboration  \cite{Belle} for fitting the rate of  $\tau\to K\pi\nu$ decay. 
The above leads to the value $31.4 \times 10^{-5}\,{\rm GeV}^{-2}$ for the vector 
form factor and $60.9 \times 10^{-6}$  
for the scalar form factor. By combining with the values $I_{+,0}$,  we obtain 
the new upper bound on the integral Eq. (\ref{eq:I}) from $\tin$ to $\infty$, 
$I_+'=(312 \pm 69) \times 10^{-5} 
{\rm GeV}^{-2} $ and  $ I_0'=(192 \pm 90)\times 10^{-6}$.
\begin{figure}[htb]
\begin{center}\vspace{0.5cm}
\vspace{0.35cm}
  \includegraphics[width=0.45 \textwidth]{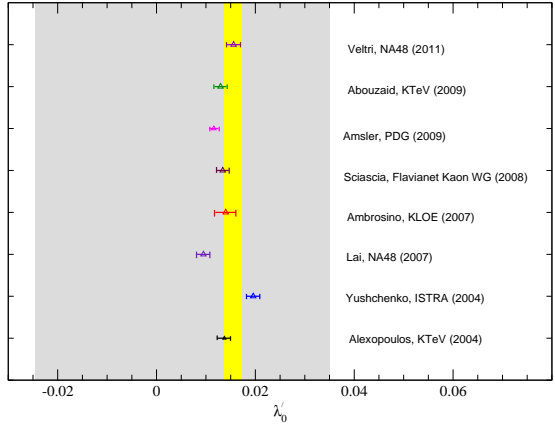}
\caption{The allowed range for the slope of the scalar form factor, when we include 
phase, modulus and the CT constraint 
(yellow band). The grey band shows the range without the CT constraint.
}
\label{fig1}
\end{center}
\end{figure}

\begin{figure}[htb]
\begin{center}
\vspace{0.35cm}
 \includegraphics[width=0.45 \textwidth]{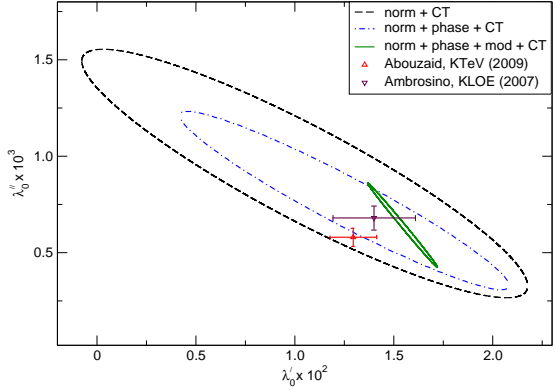}
\caption{Allowed domain for the slope and curvature of the scalar form factor, 
using the normalization $f_+(0)=0.962$, the value $f_0(\Delta_{K\pi})=1.193$,  and phase
 and modulus information up to  $\tin=(1 \, \mbox{GeV})^2$.}
	\label{fig2}
\end{center}
\end{figure}

\section{Results}

\bigskip

In Figs. \ref{fig1} and \ref{fig2}, 
the constraints for the scalar form factor are represented together with experimental 
information from various 
experiments.
As shown in Fig. \ref{fig1}, the slope $\lambda_0'$ of the scalar
form factor, predicted by NA48 (2007) 
is not consistent with our predictions (yellow band) which are obtained by taking
into account the phase, modulus as well as the CT constraint.  
Nevertheless  our 
predicted 
range for the slope is well-respected by the recent 2011 analysis by NA48 \cite{Veltri:2011zk} . 

\begin{figure}[htb]
\begin{center}
\vspace{0.35cm}
 \includegraphics[width=0.45 \textwidth]{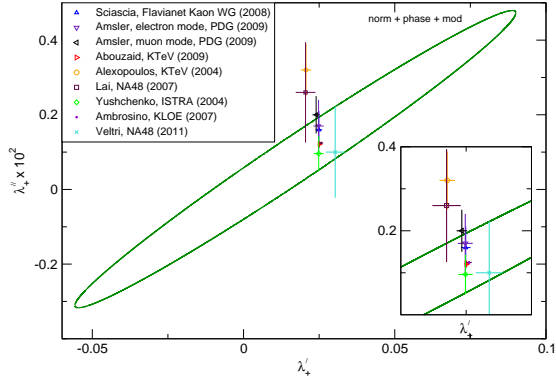}
\caption{The best constraints for the slope and curvature of the vector 
form factor in the slope-curvature plane, where the allowed region is the
interior of the ellipse.
}
	\label{fig3}
\end{center}
\end{figure}
The value of $\lambda_0'$ for this new determination 
by the 
NA48 reads 
$\lambda_0' = (15.6 \pm 1.2 \pm 0.9)\times 10^{-3}$. On the theoretical side,
the prediction of ChPT to two loops 
gives $ \lambda_0'= (13.9_{+1.3}^{-0.4}\pm0.4)\times 10^{-3}$ and  $\lambda_0''= 
(8.0_{+0.3}^{-1.7})\times 10^{-4} $ 
which are consistent with our results within errors as shown in Fig. \ref{fig2}.  
For the central value 
of the slope $ \lambda_0'$ given above, the range of  
$\lambda_0''$  is  $(8.24 \times 10^{-4}, 8.42\times 10^{-4})$. The corresponding
theoretical  
predictions are $ \lambda_0'= (16.00 \pm 1.00 )\times 10^{-3}$, $\lambda_0''= 
(6.34\pm 0.38)\times 10^{-4} $ 
obtained from dispersion relations.

\begin{figure}[ht]
\begin{minipage}[b]{0.5\linewidth}
\centering
\includegraphics[scale=0.3]{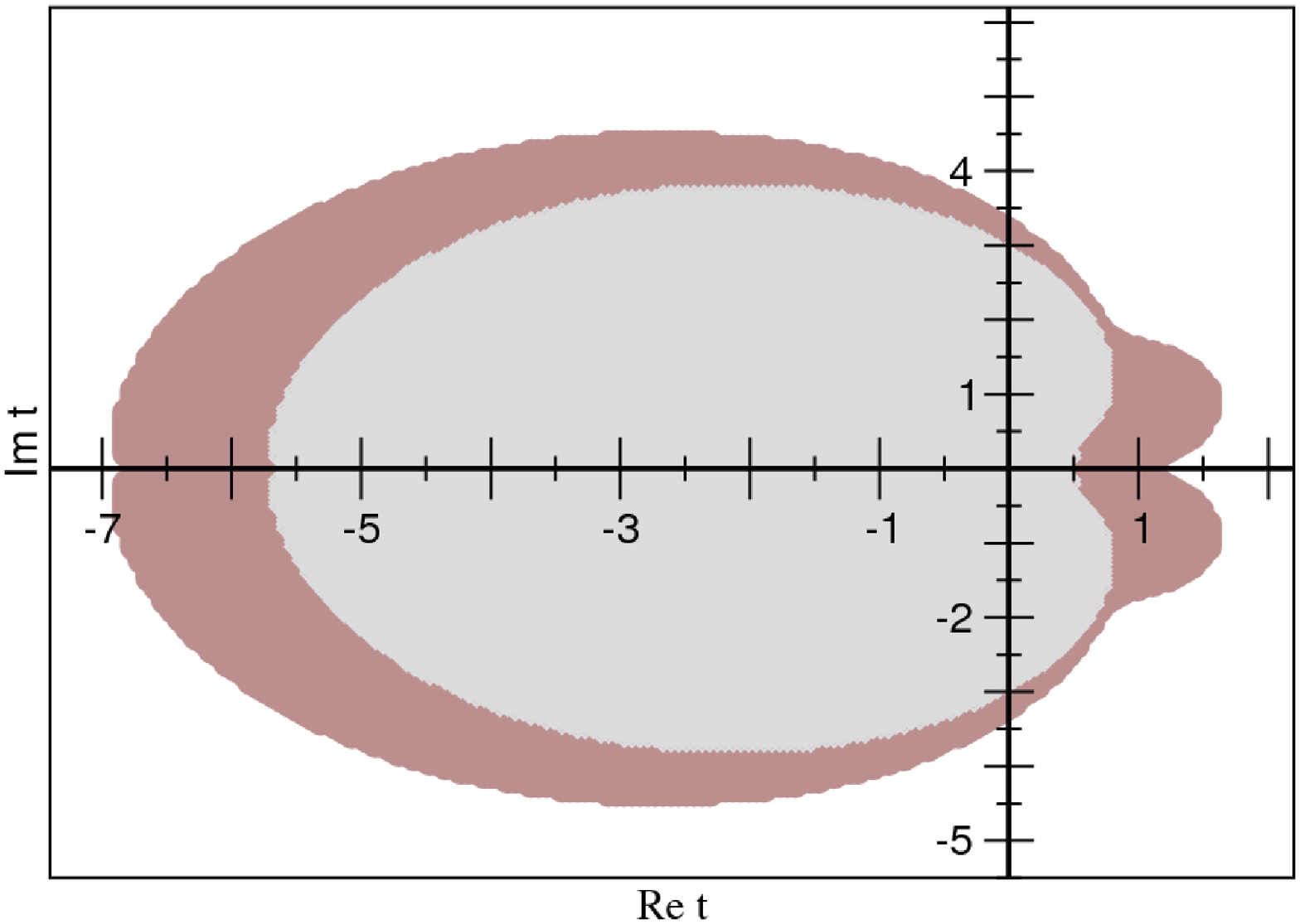}
\caption{Domain without zeros for the scalar form factor: the small domain is obtained 
without including phase and modulus in the elastic region, the bigger one using phase, modulus and CT constraint. }
\label{fig4}
\end{minipage}
\hspace{0.5cm}
\begin{minipage}[b]{0.5\linewidth}
\centering
\includegraphics[scale=0.3]{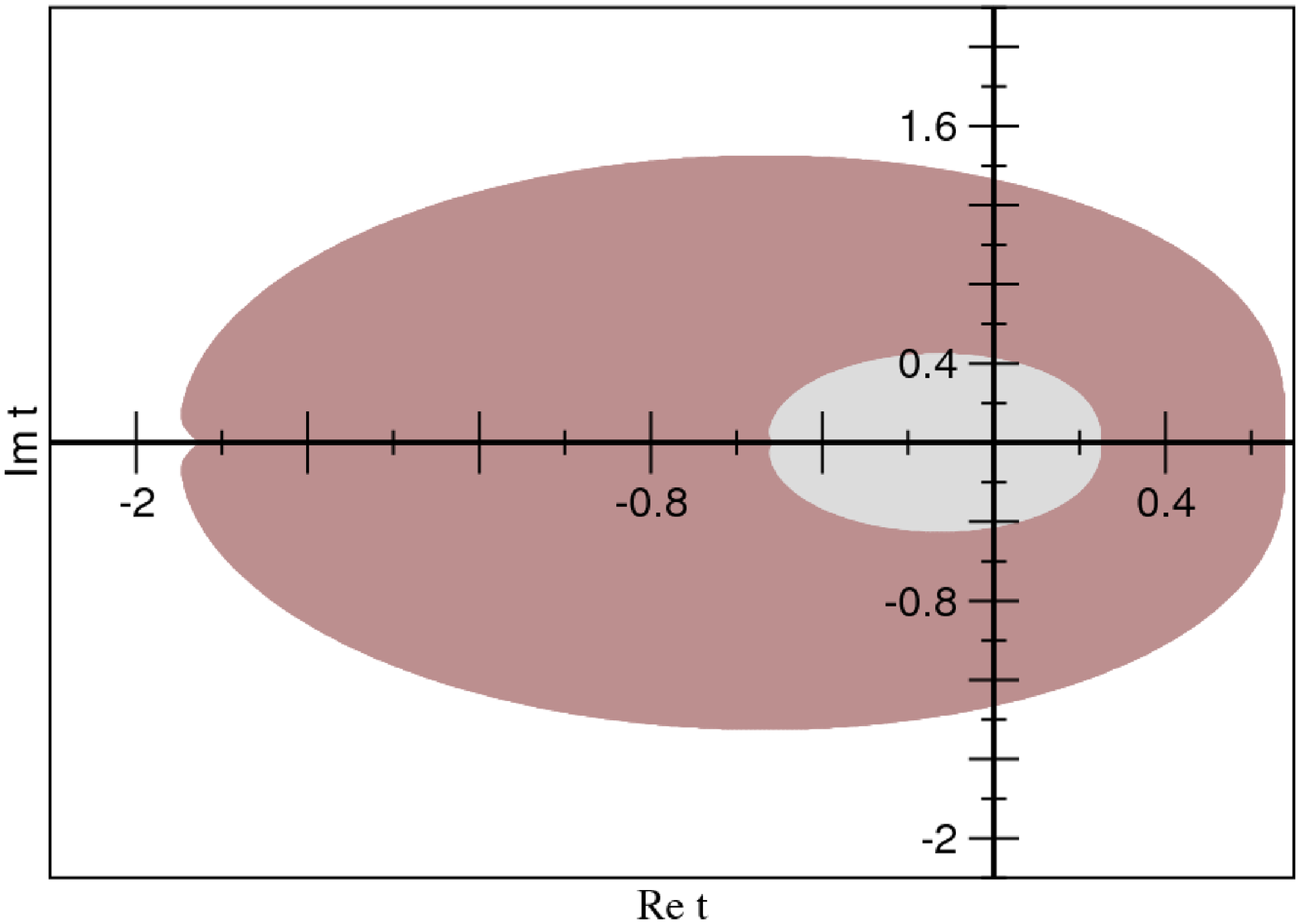}
\caption{Domain without zeros for the vector form factor: the small domain is obtained 
without including phase and modulus in the elastic region, the bigger one using 
phase and modulus.}
\label{fig5}
\end{minipage}
\end{figure}

Comparison  of the experimental results with our constraints for the vector form 
factor 
is shown in Fig. \ref{fig3}. We find that  except for the results from NA48 (2007) 
and KLOE, which have 
curvatures slightly larger than the allowed values, the experimental data satisfy 
our  constraints. 
The new results from NA48  \cite{Veltri:2011zk} provide a curvature which 
overlaps with our constraints
while the slope lies completely within our domain as seen in the figure. 
We also note  that the theoretical predictions $ \lambda_+'= (24.9 \pm 1.3)\times 
10^{-3}$, $\lambda_+''= (1.6 \pm 0.5)\times 10^{-3} $  obtained from ChPT to two loops, 
and 
$ \lambda_+'= (26.05_{-0.51}^{+0.21})\times 10^{-3}$, $\lambda_+''= (1.29_{-0.04}^{+0.01})
\times 10^{-3} $, 
and $ \lambda_+'= (25.49\pm0.31)\times 10^{-3}$, {\bf $\lambda_+''= (1.22\pm0.14)\times 
10^{-3} $ } obtained from 
dispersion relations are consistent  with the constraint. For more results, 
see \cite{Abbas:2010ns}.  

We can also extend our technique to derive regions in the complex plane where 
the form factors can not have zeros.
For the $K\pi$ form factors,
the influence of possible zeros in the context of Omn\`es
dispersive representations has been analyzed in \cite{Bernard:2009zm}.
The
absence of zeros is assumed in the recent analysis of
KTeV data reported in \cite{Abouzaid:2009ry}. 
In Fig. \ref{fig4} we show the region where zeros of the scalar form factors
 are excluded.
The formalism  rules out zeros in the physical region of the kaon semileptonic decay.
In the case of complex zeros, we have obtained a rather large region where they 
cannot be present.
In the case of the vector form factor, the analysis of \cite{Bernard:2009zm} using data from $\tau$ decay concludes that complex
zeros cannot be excluded, due to the lack of information
on the phase of the form factor in the inelastic region.
However our formalism is independent of phase information in the inelastic region
and leads without any assumptions to a rather large domain where complex zeros are excluded.
For the vector form factor,  fig. \ref{fig5} shows the region where complex zeros are 
excluded.
For more results, see \cite{Abbas:2010ns}.

\section{Conclusion}

\bigskip

We have derived stringent constraints on the shape parameters of 
the form factors of $K_{l3}$ decay which is the best source for the extraction
of CKM matrix element $V_{us}$.   
The results are promising and stringent especially in the case of the 
scalar form factor. 
The most recent 
results from NA48 
\cite{Veltri:2011zk} is consistent with our prediction for the slope of scalar 
form factor 
and restricts the range of the slope to $\sim 0.01-0.02$. We have also excluded
zeros in a rather large domain
at low energies both for the scalar and vector form factor. The Callan-Treiman 
input provides 
an additional
constraint in the case of the scalar form factor and as a result excludes a larger 
domain 
of the energy plane where zeros can exist. Thus, this work represents a
powerful application of the theory of unitarity bounds,
which relies not so much on experimental information,
but on theoretical inputs from perturbative QCD, low energy theorems and
lattice calculations.  It provides a powerful consistency 
check on determinations of
shape parameters from phenomenology and experimental analyses.


\section*{References}
\begin{thebibliography}{9}


\bibitem{Antonelli:2010yf}
  M.~Antonelli, V.~Cirigliano, G.~Isidori, F.~Mescia, M.~Moulson, H.~Neufeld, E.~Passemar and M.~Palutan {\it et al.},
  Eur.\ Phys.\ J.\ C {\bf 69}, 399 (2010). 
  [arXiv:1005.2323 [hep-ph]].
\bibitem{Cirigliano:2011ny}
  V.~Cirigliano, G.~Ecker, H.~Neufeld, A.~Pich and J.~Portoles,
  arXiv:1107.6001 [hep-ph].
\bibitem{Sher:2003fb}
  A.~Sher {\it et al.},
  Phys.\ Rev.\ Lett.\  {\bf 91}, 261802 (2003)
  [arXiv:hep-ex/0305042].
\bibitem{Ambrosino:2006gn}
  F.~Ambrosino {\it et al.}  [KLOE Collaboration],
  Phys.\ Lett.\  B {\bf 636}, 166 (2006).
  [arXiv:hep-ex/0601038].
\bibitem{Romanovsky:2007qb}
  V.~I.~Romanovsky {\it et al.},
  arXiv:0704.2052 [hep-ex].



\bibitem{Abouzaid:2009ry}
  E.~Abouzaid {\it et al.}  [KTeV collaboration],
  Phys.\ Rev.\  D {\bf 81}, 052001 (2010). 
  [arXiv:0912.1291 [hep-ex]].
\bibitem{Veltri:2011zk}
  M.~Veltri,
  arXiv:1101.5031 [hep-ex].

\bibitem{Lellouch:2009fg}
  L.~Lellouch,
  PoS LATTICE {\bf 2008}, 015 (2009). 
  [arXiv:0902.4545 [hep-lat]].

\bibitem{Durr:2010hr}
  S.~Durr {\it et al.},
  Phys.\ Rev.\  D {\bf 81}, 054507 (2010).
  [arXiv:1001.4692 [hep-lat]].

\bibitem{Colangelo:2010et}
  G.~Colangelo {\it et al.},
  Eur.\ Phys.\ J.\  C {\bf 71}, 1695 (2011).
  [arXiv:1011.4408 [hep-lat]].

\bibitem{Ramos:2011wp}
  A.~Ramos {\it et al.},
  arXiv:1101.3968 [hep-lat].

\bibitem{Abbas:2010ns}
  G.~Abbas, B.~Ananthanarayan, I.~Caprini and I.~Sentitemsu Imsong,
  Phys.\ Rev.\  D {\bf 82}, 094018 (2010). 
  [arXiv:1008.0925 [hep-ph]].

\bibitem{Abbas:2010jc}
  G.~Abbas, B.~Ananthanarayan, I.~Caprini, I.~Sentitemsu Imsong and S.~Ramanan,
  Eur.\ Phys.\ J.\  A {\bf 45}, 389 (2010).
  [arXiv:1004.4257 [hep-ph]].

\bibitem{Abbas:2009dz}
  G.~Abbas, B.~Ananthanarayan, I.~Caprini, I.~Sentitemsu Imsong and S.~Ramanan,
  Eur.\ Phys.\ J.\ A {\bf 44}, 175 (2010). 
  [arXiv:0912.2831 [hep-ph]].

\bibitem{Abbas:2011qj}
  G.~Abbas, B.~Ananthanarayan, I.~Caprini and I.~S.~Imsong,
  arXiv:1112.4270 [hep-ph].

\bibitem{anant}
  B.~Ananthanarayan, I.~Caprini, 
  {\it Constraining Form Factors with the Method of Unitarity Bounds,} these Proceedings.
  

%
 \bibitem{Baikov:2005rw}
   P.~A.~Baikov, K.~G.~Chetyrkin and J.~H.~Kuhn,
   Phys.\ Rev.\ Lett.\  {\bf 96}, 012003 (2006). 
  [hep-ph/0511063].

 \bibitem{Baikov:2008jh}
   P.~A.~Baikov, K.~G.~Chetyrkin and J.~H.~Kuhn,
   Phys.\ Rev.\ Lett.\  {\bf 101}, 012002 (2008). 
   [arXiv:0801.1821 [hep-ph]].
%
 \bibitem{Boyle:2007qe}
   P.~A.~Boyle, A.~Juttner, R.~D.~Kenway, C.~T.~Sachrajda, S.~Sasaki, A.~Soni, R.~J.~Tweedie and J.~M.~Zanotti,
   Phys.\ Rev.\ Lett.\  {\bf 100}, 141601 (2008). 
   [arXiv:0710.5136 [hep-lat]].

\bibitem{Callan:1966hu}
  C.~G.~Callan and S.~B.~Treiman,
  Phys.\ Rev.\ Lett.\  {\bf 16}, 153 (1966). 

\bibitem{Dashen:1969bh}
  R.~F.~Dashen and M.~Weinstein,
  Phys.\ Rev.\ Lett.\  {\bf 22}, 1337 (1969). 

\bibitem{Gasser:1984ux}
  J.~Gasser and H.~Leutwyler,
  Nucl.\ Phys.\ B {\bf 250}, 517 (1985).

\bibitem{Kastner:2008ch}
  A.~Kastner and H.~Neufeld,
  Eur.\ Phys.\ J.\ C {\bf 57}, 541 (2008). 
  [arXiv:0805.2222 [hep-ph]].

\bibitem{Bijnens:2003uy}
  J.~Bijnens and P.~Talavera,
  Nucl.\ Phys.\ B {\bf 669}, 341 (2003). 
  [hep-ph/0303103].

\bibitem{Bijnens:2007xa}
  J.~Bijnens and K.~Ghorbani,
  arXiv:0711.0148 [hep-ph].


\bibitem{Oehme}
R.~Oehme,
  Phys.\ Rev.\ Lett.\  {\bf 16}, 215  (1966).


\bibitem{Buettiker:2003pp}
  P.~Buettiker, S.~Descotes-Genon and B.~Moussallam,
  Eur.\ Phys.\ J.\ C {\bf 33}, 409 (2004). 
  [hep-ph/0310283].


\bibitem{ElBennich:2009da}
  B.~El-Bennich, A.~Furman, R.~Kaminski, L.~Lesniak, B.~Loiseau and B.~Moussallam,
  Phys.\ Rev.\ D {\bf 79}, 094005 (2009). 
   [Erratum-ibid.\ D {\bf 83}, 039903 (2011) ]
  [arXiv:0902.3645 [hep-ph]].


\bibitem{Moussallam:2007qc}
  B.~Moussallam,
  Eur.\ Phys.\ J.\ C {\bf 53}, 401 (2008). 
  [arXiv:0710.0548 [hep-ph]].


\bibitem{Bernard:2009zm}
  V.~Bernard, M.~Oertel, E.~Passemar and J.~Stern,
  Phys.\ Rev.\ D {\bf 80}, 034034 (2009). 
  [arXiv:0903.1654 [hep-ph]].

\bibitem{Belle}
  D.~Epifanov {\it et al.}  [Belle Collaboration],
  Phys.\ Lett.\ B {\bf 654},  65 (2007).
  [arXiv:0706.2231 [hep-ex]].

\end{thebibliography}
\end{document}